# The Dance of Atoms：De Novo Protein Design with Diffusion Model

Yujie Qin[1,2], Ming He[1,3], Changyong Yu[3], Ming Ni[1,*], Xian Liu[1,*], Xiaochen Bo[1,*]

[1]Department of Advanced & Interdisciplinary Biotechnology, Academy of Military Medical Sciences, Beijing, China

[2]School of Biological Science and Medical Engineering, Beihang University, Beijing, China

[3]College of Computer Science and Engineering, Northeastern University, Shenyang 110918, China

*Corresponding author(s). E-mail(s): niming@bmi.ac.cn, liux.bio@gmail.com, boxc@bmi.ac.cn ;
Contributing authors: qinyujie@buaa.edu.cn;
†These authors contributed equally to this work.

## Abstract

The de novo design of proteins refers to creating proteins with specific structures and functions that do not naturally exist. In recent years, the accumulation of high-quality protein structure and sequence data and technological advancements have paved the way for the successful application of generative artificial intelligence (AI) models in protein design. These models have surpassed traditional approaches that rely on fragments and bioinformatics. They have significantly enhanced the success rate of de novo protein design, and reduced experimental costs, leading to breakthroughs in the field. Among various generative AI models, diffusion models have yielded the most promising results in protein design. In the past two to three years, more than ten protein design models based on diffusion models have emerged. Among them, the representative model, RFDiffusion, has demonstrated success rates in 25 protein design tasks that far exceed those of traditional methods, and other AI-based approaches like RFjoint and hallucination. This review will systematically examine the application of diffusion models in generating protein backbones and sequences. We will explore the strengths and limitations of different models, summarize successful cases of protein design using diffusion models, and discuss future development directions.

## 1 Introduction

Proteins act as the functional executors of biological systems, performing essential tasks that sustain the complexity of life. Being composed of 20 amino acids, proteins can form an immense number of combinations. Natural evolution has explored only a tiny fraction of the "protein space". Expanding this space by designing entirely novel proteins—not found in nature—offers a promising strategy to satisfy human needs and address global challenges.

In the early stages of de novo protein design, researchers relied on computational methods guided by basic physical and chemical principles[1]. While these approaches successfully designed water-soluble proteins, they struggled with larger, cooperatively folded proteins rich in β-sheets[2]. As proteins in structure databases expanded[3–5] and computational power advanced, fragment-based methods emerged as a mainstream approach. These methods assemble naturally occurring protein backbone fragments into new structures. Over the past decade, researchers have successfully designed proteins for immune signaling, targeted therapies, and other applications. Despite their achievements, fragment-based methods had a low success rate and required much computation and experimentation.

In recent years, advances in AI have revolutionized protein design. Generative AI models can extract patterns from large datasets and parameterize complex relationships, offering a powerful alternative to traditional methods. Recent studies have applied various generative models to protein design[6–8]. Among these, diffusion models stand out for their ability to learn complex distributions, handle high-dimensional data, and generate diverse outputs[9]. Over the past two to three years, more than ten diffusion-based protein design models have emerged, achieving far higher success rates than traditional methods.

This review provides an overview of the application of diffusion models in protein design, especially in structure generation. We compare the strengths and limitations of these methods and evaluate model performance from multiple perspectives: efficiency, structure plausibility, designability, naturalness, diversity, and novelty. With these evaluation criteria, we aim to guide researchers in selecting suitable tools. Additionally, we discuss the advantages, challenges, and future opportunities of applying diffusion models to *de novo* protein design.

## 2 The Fundation of Diffusion Models

Diffusion models have revolutionized generative AI. Their fundamental concept involves corrupting data by noise, and then training a neural network to denoise it. These models can generate outputs resembling the training data distribution from noise. Compared to other generative AI models, such as Generative Adversarial Networks (GANs) and Variational Autoencoders (VAEs), diffusion models excel at modeling complex data distributions, resulting in more diverse and accurate outputs. Furthermore, by conditioning the model with specific requirements during the training process, it can generate samples tailored to particular needs.

Originally proposed by Sohl-Dickstein et al., diffusion models have advanced significantly. In 2019, Song et al. introduced the Noise-Conditional Score Network (NCSN), which estimates the "Stein score" under different noise levels[10]. However, estimation of the Stein score and sampling based on Langevin dynamics make the training and generation process of NCSN less intuitive. Ho et al. simplified this process further by proposing Denoising Diffusion Probabilistic Models (DDPM)[11], which predict and iteratively remove noise, proving diffusion models' competitiveness in image generation. Because the backward process of DDPM uses a Markovian process that involves thousands of steps, it is time-consuming. To accelerate this, Jiaming Song et al. introduced Denoising Diffusion Implicit Models (DDIM) with a non-Markovian process[12], while Yang Song et al. linked stochastic and ordinary differential equations (SDEs/ODEs)[13], reducing the steps needed from thousands to a few dozen.

Diffusion models have continually improved, leading to significant enhancements in tasks like image and even video generation. These advancements lay a strong foundation for applying diffusion models in more complex scientific fields, such as de novo protein design.

## 3 Protein Design with Diffusion Models

De novo protein design has two main approaches: one designs the protein structure first and then determines the amino acid sequence, while the other directly designs the sequence. With the introduction of diffusion models, three types have emerged: backbone structure generation models, sequence generation models, and structure and sequence co-generation models (Fig. 1). Backbone structure generation models apply the diffusion process to the atomic spatial positions of the protein backbone, leveraging 3D information and excelling in functional protein design. They are the mainstream approaches but require a reverse folding process to derive the amino acid sequence. Sequence generation models perform diffusion directly on the protein's amino acid sequence. They are generally more straightforward and efficient than backbone structure generation models. Besides, they have a larger training dataset, consisting of over 200 million sequences in contrast to fewer than 220,000 structures. However, they struggle with function-related tasks like protein binder design and scaffold design. To overcome these limitations, some models perform diffusion on structure and sequence simultaneously, referred to as structure and sequence co-generation models.

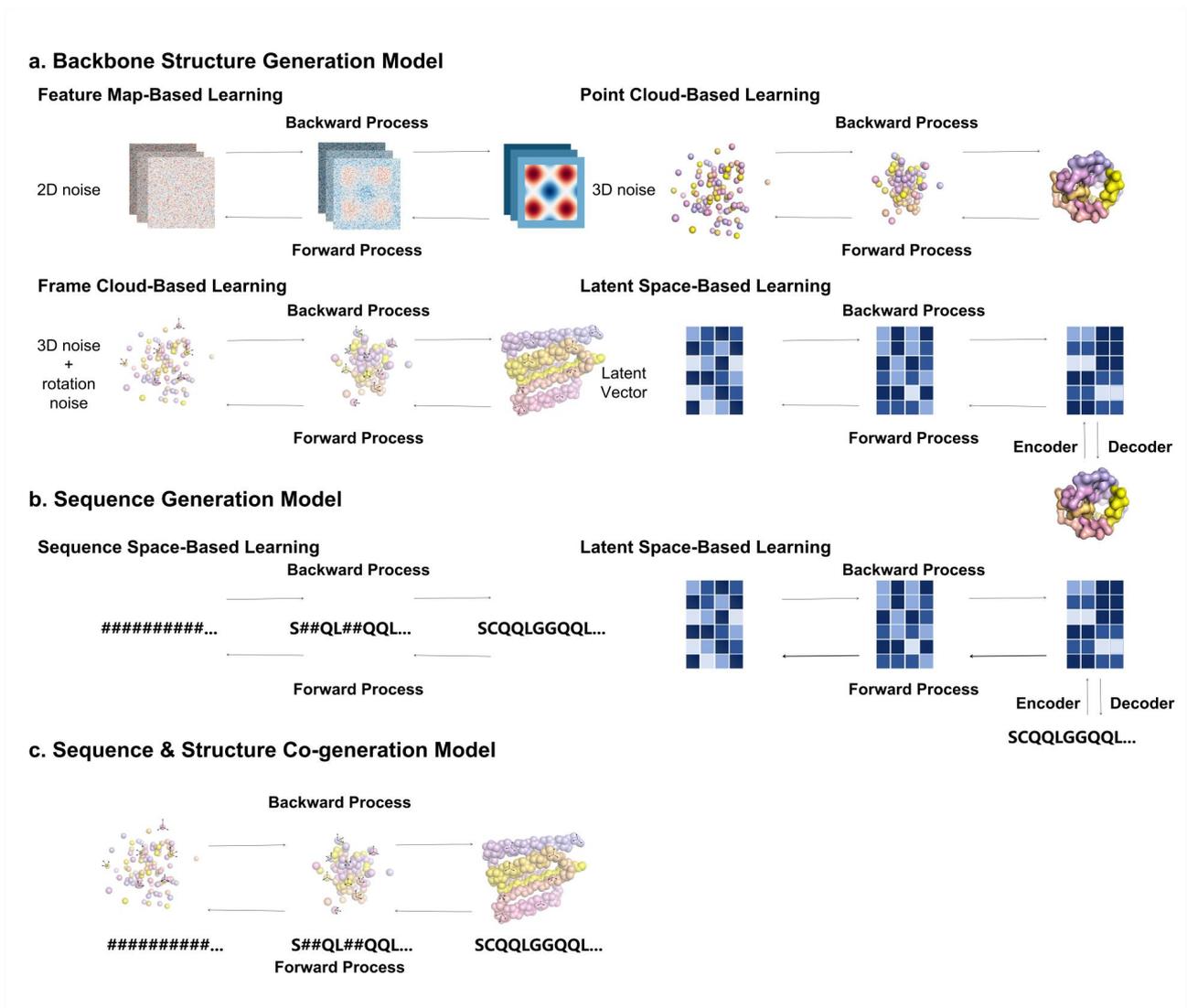

Fig.1 Classification of de novo protein design tools using diffusion models: a. Backbone structure generation model. b. Sequence generation model. c.Sequence and structure co-generation model.

Table 1  Work on  Protein Design with diffusion models

| Model | Type | Frame | Noise Type | Protein representation | SE(3) Equivariance & SE(3) Invariance | Length Range |
|---|---|---|---|---|---|---|
| Genie | Backbone Structure Generation | DDPM | Isotropic 3D Gaussian noise | Point cloud & Frame cloud | SE(3) Equivariance | 0-256 |
| Chroma | Backbone Structure Generation | DDPM | Isotropic 3D Gaussian noise | Point cloud | ChromaBackbone: SE(3) Equivariance; ChromaDesign: SE(3) Invariance | No limitation |
| ProtDiff | Backbone Structure Generation | DDPM | Isotropic 3D Gaussian noise | Point cloud | SE(3) Equivariance | No limitation |
| ProteinSGM | Backbone Structure Generation | SGM | Isotropic 2D Gaussian noise | Feature Map | SE(3) Invariance | 0-128 |
| Foldingdiff | Backbone Structure Generation | DDPM | Wrapped Gaussian | Frame cloud | SE(3) Invariance | 0-128 |
| FrameDiff | Backbone Structure Generation | SGM | Isotropic 3D Gaussian noise & discretized Brownian motion | Frame cloud | SE(3) Invariance | No limitation |
| SCUBA-D | Backbone Structure Generation | DDPM | mask，isotropic 3D Gaussian noise，rotation noise | Feature Map | SE(3) Equivariance | No limitation |
| PVQD | Backbone Structure Generation | DDPM | Isotropic 2D Gaussian noise | Latent space vector | SE(3) Invariance | No limitation |
| RFdiffusion | Backbone Structure Generation | DDPM | Isotropic 3D Gaussian noise & discretized Brownian motion | Frame cloud | SE(3) Invariance | No limitation |
| Evodiff | Sequence Generation | DDPM | mask | Amino acid sequence | / | No limitation |
| TaxDiff | Sequence Generation | DDPM | Isotropic 2D Gaussian noise | Amino acid sequence | / | Unable to Control Length |
| full-atom DDPM | Sequence & Structure Co-generation Model | DDPM | mask, slerp, linear interpolation | Frame cloud | SE(3) Equivariance | No limitation |

## 3.1 Backbone Structure Generation Model

The protein backbone forms the primary structure of a protein and typically consists of four atoms per residue: amino nitrogen, carboxyl oxygen, Cα, and carboxyl carbon. The backbone is crucial in determining the protein's overall structure and stability. Designing the backbone is more efficient and requires fewer computational resources than conducting a full-atom design. Thus, many studies employ diffusion models to design the backbone and subsequently generate the amino acid sequence based on it.

Since proteins can be represented in various forms, selecting the right representation is the initial step in backbone design[14]. Several validated representation methods have been established, including feature maps, point clouds, reference frame clouds, and latent space vectors (Fig. 1a). This section will explore these four methods, discussing their characteristics, advantages, limitations, and applicability in different scenarios.

### 3.1.1 Feature Map-Based Learning

Protein structures can be transformed into two-dimensional feature maps by extracting structural features such as bond angles, interatomic distances, and others. The feature map method can quickly adapt to the mature frameworks developed for image generation, as seen in ProteinSGM by Lee et al.[15] and SCUBA-D by Liu et al.[16].

The difference between the two models lies in how they represent protein structures. ProteinSGM maps the protein structure onto four feature maps, including Cβ-Cβ distances of two adjacent residues, the torsion angles (ω, θ), and the Φ plane angle. SCUBA-D, however, represents relative translations and rotations between residues. Additionally, ProteinSGM employs a score-based diffusion model using RefineNet to predict the Stein score, while SCUBA-D uses a DDPM framework with a deep neural network to predict image noise. ProteinSGM outputs feature maps that require post-processing (MinMover, FastDesign, FastRelax) to obtain backbone coordinates, whereas SCUBA-D directly generates backbone atom coordinates. In the forward process, ProteinSGM designs its noise as follows:

$$\delta x = \sqrt{\frac{\delta[\sigma^2(t)]}{\delta t}} \delta w \qquad (1)$$

Where $\delta w$ represents the standard Wiener process, and $\sigma(t) = \sigma_{min}(\frac{\sigma_{max}}{\sigma_{min}})^t$, with $\sigma_{max}$ and $\sigma_{min}$ being user-defined noise variance boundaries. SCUBA-D employs adversarial loss with a GAN discriminator to evaluate predictions and features two consecutive denoising modules. First, a single-step denoising module generates the prior backbone, followed by a high-resolution diffusion module that refines it, assisted by a pre-trained language model for amino acid sequences. ProteinSGM, in contrast, is not an end-to-end model. It first generates images, which can be converted into 3D structures through multiple post-processing steps. These steps increase complexity, prolong generation time, and introduce noise during the post-processing stage, resulting in less realistic generated structures[17]. Moreover, due to the inherent characteristics of image-based data, ProteinSGM is limited to proteins with up to 128 amino acids. In contrast, SCUBA-D does not impose such restrictions and can directly generate protein backbones, simplifying the design process.

### 3.1.2 Point Cloud-Based Learning

Point clouds represent protein structures using 3D coordinates, offering a more intuitive approach than 2D feature maps. Both ProtDiff[18] and Genie[19] perform diffusion on point cloud data, specifically the 3D coordinates of Cα atoms, resulting in backbone structures containing only these coordinates.

Although both models use point cloud representations, ProtDiff models proteins as fully connected graphs, while Genie directly diffuses Cα coordinates. ProtDiff follows a linear noise-adding strategy, whereas Genie iteratively adds Gaussian noise using cosine variance. Additionally, while Genie utilizes an SE(3)-equivariant network, ProtDiff provides greater equivariance by employing an E(3)-equivariant GNN for denoising. Regarding output, ProtDiff can generate proteins with over 500 residues, whereas Genie is limited to 256. ProtDiff also supports motif-constrained scaffold design through SMCDiff, while Genie is limited to unconditional generation.

Both ProDiff and Genie perform diffusion on protein backbones to reduce memory consumption and simplify noise-adding. However, this approach ignores the orientation between residues. Additionally, designing suitable sequences based solely on Cα coordinates remains a challenge.

### 3.1.3 Frame Cloud-Based Learning

The reference frame cloud method represents each residue as a freely floating backbone reference frame, where a protein's 3D backbone structure with N residues is modeled as N local reference frames, each rotating and translating independently relative to a global reference frame. This approach was one of the key factors to AlphaFold2's success. Specifically, the method represents the protein backbone as a graph, where each residue's position and rotation matrix are determined using its Cα atom as the origin. The rotation matrix is constructed with the vector from Cα to C as a basis vector, and atomic coordinates are orthogonalized using the Gram-Schmidt process. Unlike the point cloud method, the reference frame cloud retains angular information between residues, enabling the interpretation of chemical chirality[20]. The additionally structural information makes it particularly useful for accurately modeling residue interactions and orientations.

Diffusion on reference frame clouds has been applied in models such as FrameDiff[21], RFdiffusion[8], and Chroma[22]. FoldingDiff[17] further simplified the method by omitting atomic coordinates, treating the backbone as a sequence of six internal consecutive angles.. FrameDiff and RFdiffusion apply isotropic Gaussian noise separately to the translational and rotational components of reference frames. The noise addition process for these models can be described as follows:

$$\delta X^{(t)} = [0, -\frac{1}{2} X^{(t)}] dt + [dB^{(t)}_{SO(3)}, dB^{(t)}_{\mathbb{R}^3}] \quad (2)$$

Where $dX^{(t)}$ represents the added noise, $X^{(t)}$ denotes the sample at time t, $dB^{(t)}_{SO(3)}$ represents Brownian motion in the SO(3) space, and $dB^{(t)}_{\mathbb{R}^3}$ represents Brownian motion in the $\mathbb{R}^3$ space. Isotropic noise forces the model to learn correlations in the data from scratch. In contrast, Chroma uses anisotropic noise, introducing chain and gyration radius constraints into the diffusion process, which gradually transforms the protein structure into a randomly folded polymer. Chroma's noise addition scheme includes two parameters and follows the Log-linear SNR strategy[23]. The noise

addition process can be described as:

$$\delta X^{(t)} = -\frac{1}{1-t} X^{(t)} dt + \sqrt{\frac{2t}{1-t}} R dw_t \qquad (3)$$

Where, R represents the square root of the covariance matrix. Besides, the FoldingDiff model samples noise from a wrapped normal distribution, which can be expressed as:

$$N_{wrapped}(x_t; \sqrt{1-\beta_t} x_{t-1}, \beta_t I) \qquad (4)$$

Where $\beta_t \in (0,1)_{t=1}^T$ is determined by the cosine variance table[24].

The loss functions for the four models are set differently. FrameDiff's loss includes $L_{dsm}$ (translation and rotation errors of residues), $L_{bb}$ (global error of the atoms), and $L_{2D}$ (local error of the atoms), with $L_{bb}$ and $L_{2D}$ used as auxiliary loss during early diffusion steps. RFDiffusion's loss, in addition to the elements in FrameDiff's loss, also includes errors in dihedral and planar angles. Chroma train model by optimizing a bound on the log marginal likelihood of data, also called Evidence Lower Bound (ELBO), with optional auxiliary functions. FoldingDiff divides its loss into L1 loss for high error and L2 loss for low error. The network frameworks also differ among the four models. FrameDiff, RFDiffusion, and Chroma use the SGM training strategy, with FrameDiff fine-tuned with FramePred and RFDiffusion fine-tuned on RosettaFold. FoldingDiff uses the DDPM strategy.

These four models each possess distinct strengths and weaknesses. In terms of the methodology, FrameDiff and RFDiffusion utilize a diffusion method for orientation-preserving rigid motion space SE(3) in $R^3$. In contrast, FoldingDiff employs a simplified backbone representation to model protein folding into stable conformations, eliminating the necessity for complex equivariant networks. However, FoldingDiff's output can be significantly influenced by the denoising errors in the early stages. Regarding their capabilities, RFDiffusion and Chroma can design protein structures for various tasks, while the other two models primarily generate structures randomly. Additionally, FrameDiff, RFDiffusion, and FoldingDiff are limited to generating protein backbone, with Foldingdiff restricting the residue length to 40-128 amino acids. In contrast, Chroma achieves full-atom modeling by incorporating a structure-based sequence generation module alongside the protein backbone design module.

### 3.1.4 Latent Space-Based Learning

The latent space vector method encodes protein structures into a low-dimensional vector representation, reducing the complexity of handling high-dimensional space while preserving structural diversity and design potential. The PVQD model[25] performs diffusion in the latent space vectors of protein structure mappings. During training, PVQD uses a graph-based geometric vector perceptron (GVP) to encode the protein structure, and the encoded vector represents the latent space. Each vector is replaced by the nearest quantized vector from a codebook, and the protein backbone is generated through a decoder. In the generation phase, PVQD samples directly from the latent space, replacing the sampled vector with a quantized one, and generates a new protein backbone. This method smooths the data distribution by transforming proteins from a complex 3D space to a low-dimensional latent space, allowing diverse data generation without handling original space details.

## 3.2 Sequence Generation Model

Structure generation protein design has two major limitations. Primarily, protein structures are not always static, making this approach ineffective for proteins with dynamic properties. The most extreme example is intrinsically disordered regions. Besides, limited structural data restricts its ability to capture native sequence diversity—while the PDB holds about 200,000 resolved structures, billions of unique native sequences exist. Sequence generation models may overcome these issues, leading researchers to integrate diffusion models into sequence-based design.

Diffusion models for protein sequence design are mainly divided into two approaches: directly generating protein sequences and generating protein sequences from latent space (Fig. 1b). We will introduce key research in these methods, analyze their pros and cons.

### 3.2.1 Sequence Space-Based Learning

Recent sequence generation models operate in sequence space, using discrete diffusion to mask and reconstruct sequences progressively. Notable models include those by Bo Ni et al.[26], Yang et al.[27], EvoDiff[28], and TaxDiff[29]. Although with similar diffusion process, these models are different in constraint condition. Bo Ni et al.'s model and Yang et al.'s model focus on sequence generation under specific structural constraints. However, with a key difference: Bo Ni et al. use secondary structure information as a condition, whereas Yang et al.'s model relies on 3D backbone structures for control. EvoDiff incorporates evolutionary-scale data, using multiple sequence alignment (MSA) to capture conservation and variation patterns in amino acid sequences. TaxDiff introduces taxonomic information as a condition, enhancing the species specificity of the generated protein sequences.

### 3.2.2 Latent Space-Based Learning

While discrete diffusion on sequences is effective, there is still room for improvement in the quality of the generated sequence. Similar to structure generation models, sequence generation models can utilize latent-space diffusion. Specifically, during the forward process, amino acid sequences are mapped into a latent space via an encoder and gradually diffused into a noise distribution. In the backward process, the model samples from the noise distribution, denoises it into latent vectors, and then maps these vectors back into amino acid sequences using a decoder. For example, the DiMA model developed by Meshchaninov et al. employs the ESM-2 pLM as an encoder to obtain continuous representations of protein sequences[30].

Sequence generation diffusion models offer significant advantages. Sequence data is lighter than structure data, enabling faster training and inference. Moreover, the vast availability of sequence data provides a richer training set. The success of AlphaFold2 also highlights that sequences encode structural information. However, current models remain largely stochastic, lacking robust methods for designing proteins with specific functions, which limits their practical applications.

## 3.3 Sequence & Structure Co-generation Model

Conventional methods for generating backbone structures or sequences often lead to the loss of natural protein information in training data, resulting in suboptimal data utilization. To address this issue, Namrata Anand et al. proposed a diffusion-based co-generation approach that simultaneously designs both the protein backbone and side chains, improving efficiency and design quality[31] (Fig. 1c).

This diffusion model enables protein co-generation by learning the 3D coordinates of the backbone, global rotation represented by unit quaternions, residue amino acid types, and the torsion angles of side chains attached to the main-chain atoms. However, during the noise-adding process, rotational variables and the four χ angles do not lie on an Euclidean manifold, making it impossible to apply simple random scaling and perturbation for diffusion. To overcome this, the researchers first sample from a uniform random rotation and then apply spherical linear and unit interpolation to interpolate the sample smoothly. This approach ensures a well-defined diffusion process. The model can generate protein samples that satisfy structural constraints when provided with secondary structure information and coarse constraints, such as the Cα distance matrix.

While its generative performance still needs improvement, the simultaneous co-generation of sequence and structure offers a promising direction for the future development of protein design models.

# 4 Criteria for evaluating the performance of protein design models

We evaluated eight publicly available diffusion model-based structure generation models across six key metrics: efficiency, structural plausibility, designability, novelty, naturalness, and diversity (Fig. 2). Efficiency is measured by the computational time taken for each structure, while structural plausibility is evaluated through the distributions of dihedral angles. To assess designability, we look at self-consistency TM-scores (sc-TM), and novelty is determined by how far the data deviates from the training set. Naturalness is judged based on the distributions of secondary structures, and diversity is gauged through structural clustering

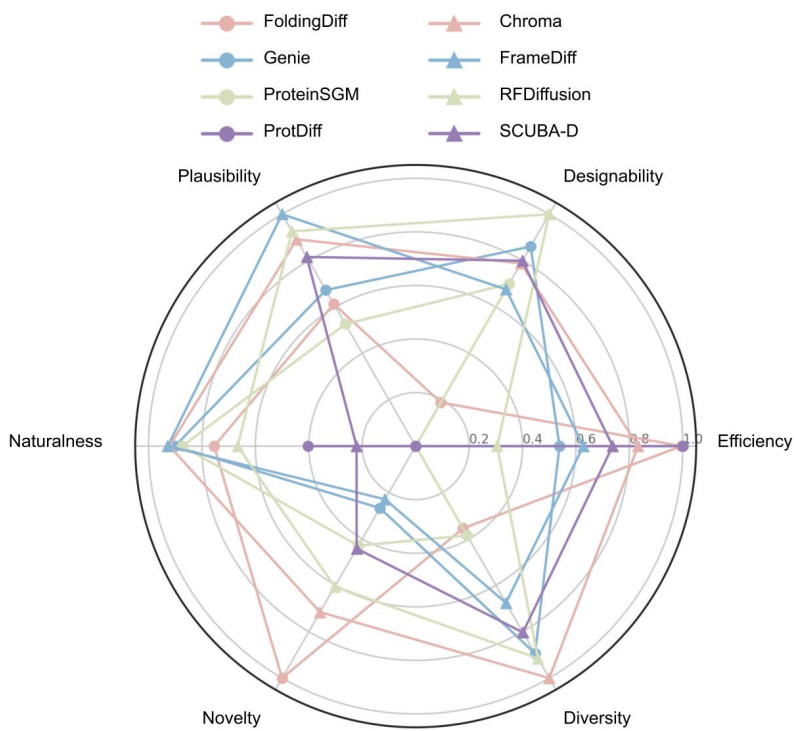

Fig. 2 Comprehensive Evaluation of Diffusion Model-Based De Novo Protein Design Frameworks

The results indicate that RFdiffusion and Chroma exhibit the most balanced performance, while other models excel in specific aspects. FrameDiff demonstrates superior structural plausibility while retaining a commendable level of naturalness. In contrast, RFDiffusion is distinguished by its designability, while Chroma excels in diversity while still upholding a strong sense of naturalness. Additionally, FoldingDiff displays significant novelty coupled with efficient performance (Fig. 2). Notably, FoldingDiff shows a high novelty but relatively low diversity, which may stem from its model architecture and training data selection. Similarly, ProteinSGM shows a similar trend. Despite employing similar training strategies, FrameDiff and RFDiffusion yield different results across various metrics. This indicates that incorporating pre-trained structure prediction models could improve designability, diversity, and novelty in protein design.

## 4.1 Efficiency

For each model, we generated 100 single-chain protein structures under unconditional settings at six residue lengths: 50, 100, 150, 200, 300, and 500. All structures were generated using an NVIDIA A100 GPU, and the average computational time per model was recorded. ProtDiff required the shortest runtime among the evaluated models, while ProteinSGM required the longest. Except for Genie, these models' computational time increases as the residue length grows. It's also worth noting that some models have specific limitations. For example, FoldingDiff and ProteinSGM can only generate proteins within the range of 0 to 128 residues, Genie-SwissProt supports sequences of up to 256 residues.

Although all models claim to generate protein backbone structures, the atomic-level outputs vary (Table 2). Chroma follows an integrated pipeline that generats both structural and sequence information, providing full-atom coordinates. ProteinSGM outputs backbone atoms including amide nitrogen (N), amide hydrogen (H), alpha carbon (Cα), carbonyl carbon (C), carbonyl oxygen (O), and side-chain beta carbon (Cβ). FrameDiff provides a similar set but omits the amide hydrogen. RFdiffusion and SCUBA-D further exclude side-chain carbon atoms, while FoldingDiff additionally omits the carbonyl oxygen. ProtDiff and Genie output only Cα coordinates. All subsequent evaluations are based on the structures generated in this stage.

Table 2 Atomic Types Included in Model-Generated Structures

| Molde | Cα | N | H | C | O | Cβ | Side-chain Atoms |
|---|---|---|---|---|---|---|---|
| Chroma | √ | √ | √ | √ | √ | √ | √ |
| ProteinSGM | √ | √ | √ | √ | √ | √ | × |
| FrameDiff | √ | √ | × | √ | √ | √ | × |
| RFdiffusion | √ | √ | × | √ | √ | × | × |
| SCUB-D | √ | √ | × | √ | √ | × | × |
| FoldingDiff | √ | √ | × | √ | × | × | × |
| ProtDiff | √ | × | × | × | × | × | × |
| Genie | √ | × | × | × | × | × | × |
| PVQD | √ | √ | × | √ | √ | × | × |

## 4.2 Structural Plausibility and Designability

Despite diffusion models can generate various protein structures, whether these structures can stably exist and perform proper function in the real world remains a significant concern. Given that biological experiments are time-consuming and labor-intensive, we assess the model's output from two in-silico criterias——structural plausibility and designability. This evaluation lays the groundwork for furthersequence design, and experimental validation.

The structural plausibility of proteins refers to whether the generated protein structures meet native proteins' statistical patterns and stability requirements in terms of geometry and physical properties. We evaluate the structural plausibility of the generated structures by analyzing the distribution of dihedral angles. Dihedral angles are key parameters that describe the rotational angles between backbone atoms in protein polypeptide chains. We primarily focus on the rotational angles of the N–Cα bond ($\varphi$) and the Cα–C bond ($\psi$). Since FoldingDiff, Genie, and ProtDiff generate only Cα atom coordinates, directly obtaining dihedral angle distributions of these models' outputs is not feasible. To address this, we utilize ProteinMPNN to predict the amino acid sequence of the structure, then employ Omegafold to predict the corresponding structure for that sequence and calculate the TM-score between the two structures, and finally select the highest-scoring prediction as our evaluation targets.

To obtain the dihedral angle distribution in native structures, we extracted all protein single chains from experimentally determined structures in the PDB database. After deduplication using MMseq clustering, we selected protein chains with lengths of 25–75, 75–125, 125–175, 175–225, 275–325, and 475–525 for further evaluation, corresponding to generated lengths of 50, 100, 150, 200, 300, and 500, respectively. We project all the dihedral angles of model-generated or native structures onto the $\psi$-$\varphi$ plane and visualize them based on their distribution density. To enhance visualization, we first applied a logarithmic transformation to normalize the results within a smaller range. Additionally, to eliminate brightness differences between different tools, we scaled the results using the ratio of the total number of dihedral angles in native proteins to that in generated structures (Figure 3a). Furthermore, to quantify the distributional differences in dihedral angles between generated and natural structures, we calculated the mean squared error (MSE) between them (Figure 3b). The results indicate that the dihedral distribution generated by FrameDiff closely resembles that of native proteins, while ProtDiff shows a larger discrepancy. Excluding models that generate only Cα atom coordinates, ProteinSGM demonstrates the greatest deviation from the native dihedral angle distribution.

Designability focuses on whether amino acid sequences exist that can fold to the generated structures. Typically assessed using the self-consistent TM-score method proposed by Trippe et al. This method first designs the sequences of the designed structure then predicts the structure of the designed sequences and finally computes the TM-score between the designed structure and the predicted structures (sc-TM). A sc-TM value closer to 1 indicates higher consistency between the generated and predicted structures. High designability suggests that a structure can support reasonable amino acid sequences and has the potential to yield stable, functional proteins upon experimental validation.

In the evaluation process, we first applied two structure-based sequence design models, ProteinMPNN[32] and CarbonDesign[33], to backfold the generated structures. Specifically, ProteinMPNN was set with a sampling temperature of 0.1, generating eight sequences per input structure. Next, we used OmegaFold[34] to predict the three-dimensional structures of the designed sequences and assessed the TM-score or RMSD between the generated and predicted structures. To minimize bias introduced by structure prediction models, we analyzed the confidence scores (pLDDT values) of the predicted structures and examined the correlation between sc-TM and pLDDT (Figure 3c). Since different models may perform optimally at different sequence lengths, we further evaluated designability across various length conditions (Figure 3b). The results show that RFDiffusion maintains the highest structural consistency across all lengths, while ProtDiff exhibits the lowest self-consistency at all lengths except for 100 residues. When generating structures with 100 residues, FoldingDiff has the lowest structural self-consistency.

To ensure that subsequent evaluations are based on designable protein structures, we set a threshold for filtering generated structures using the 5th percentile of the sc-TM distribution from native proteins. Specifically, 95% of native protein structures have a sc-TM above this threshold, and only generated structures with a sc-TM exceeding this value are selected for further evaluation.

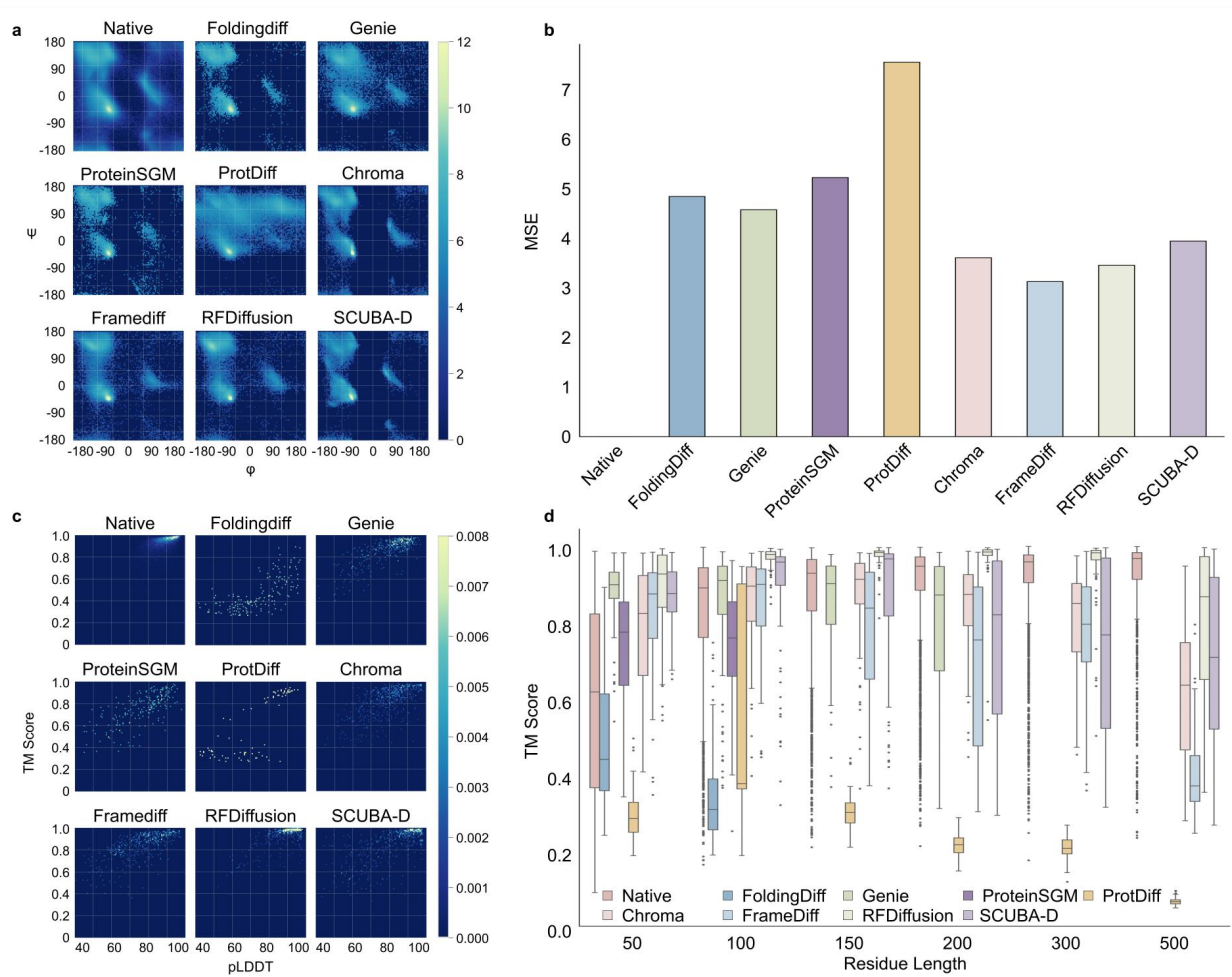

Fig. 3 a. Dihedral angle distribution of native and generated proteins. b. Mean squared error of dihedral angle distribution between generated and native proteins. c. Relationship between the sc-TM and the pLDDT for native and generated proteins. d. Distribution of sc-TM for native and generated proteins across different lengths.

## 4.3 Novelty

Novelty is a key metric for assessing the creativity of generative models, as it gauges how different the generated results are from the training data. In protein design, novelty indicates a model's ability to produce innovative structures that are significantly different from previously known protein structures. We use TM-score and RMSD—both indicators of protein structural similarity—to quantify the distance between generated results and existing structures[35]. The TM-score emphasizes global structural similarity, where lower values indicate greater differences. RMSD focuses on local atomic deviations, where higher values signify greater dissimilarity[36]. Specifically, for each protein structure generated by the model, we compute its TM-score and RMSD against all protein chains in the training set. This allows us to identify the most similar structure from the training data. The minimum TM-score and maximum RMSD are then used as novelty metrics to represent the greatest possible difference between generated results and known structures. Additionally, to comprehensively evaluate the model's creative capability, we further analyze the distribution of TM-scores and RMSD values across different lengths (Fig. 4a, b).

In terms of global structure (Fig. 4a) FoldingDiff exhibits highest novelty when generating short protein chains. For medium-length protein chains, SCUBA-D, Chroma, and RFDiffusion all demonstrate strong creativity. For longer protein chains, SCUBA-D outperforms the other models. Regarding local structural differences (Fig. 4b), reflect similar trends as the TM-score, and ProtDiff shows the highest creativity when generating protein chains of length 50.

Table 3 Training Set of Model

| Model | Training set |
|---|---|
| RFdiffusion | PDB |
| Chroma | PDB |
| Foldingdiff | CATH 4.3.0 |
| Genie | Scope dataset |
| ProtDiff | PDB |
| FrameDiff | PDB |
| ProteinSGM | CATH 4.3.0 |
| SCUBA-D | CATH 4.2.0 |
| PVQD | PDB |

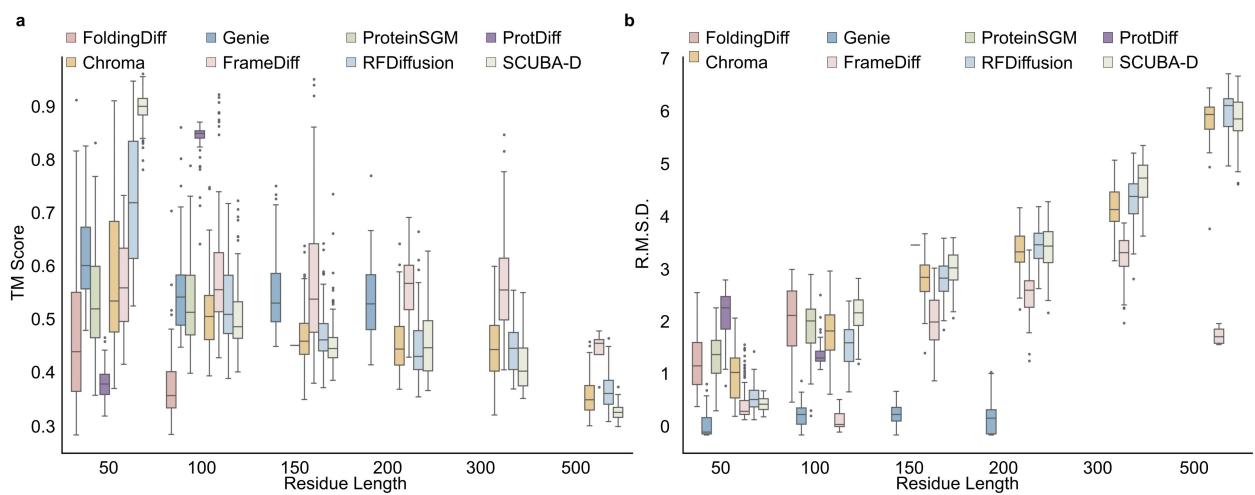

Fig. 4 a. Maximum TM-score distribution between generated proteins and their closest training set counterparts. b. Minimum RMSD distribution between generated proteins and their closest training set counterparts.

## 4.4 Naturalness

The "naturalness" of protein structures measures the similarity between model-generated and native proteins. To evaluate this, we first analyze the secondary structure composition (α-helices, β-sheets, and loop regions) of generated protein using the DSSP algorithm and quantify the distributional characteristics of these structural elements (Fig. 5a). Then we employ the Kullback-Leibler (KL) divergence to measure the differences between the distributions of generated and native protein structures (Fig. 5b). Specifically, we partition the distributions of secondary structures into 50 equally spaced bins, normalize them into probability distributions, and compute the KL divergence.

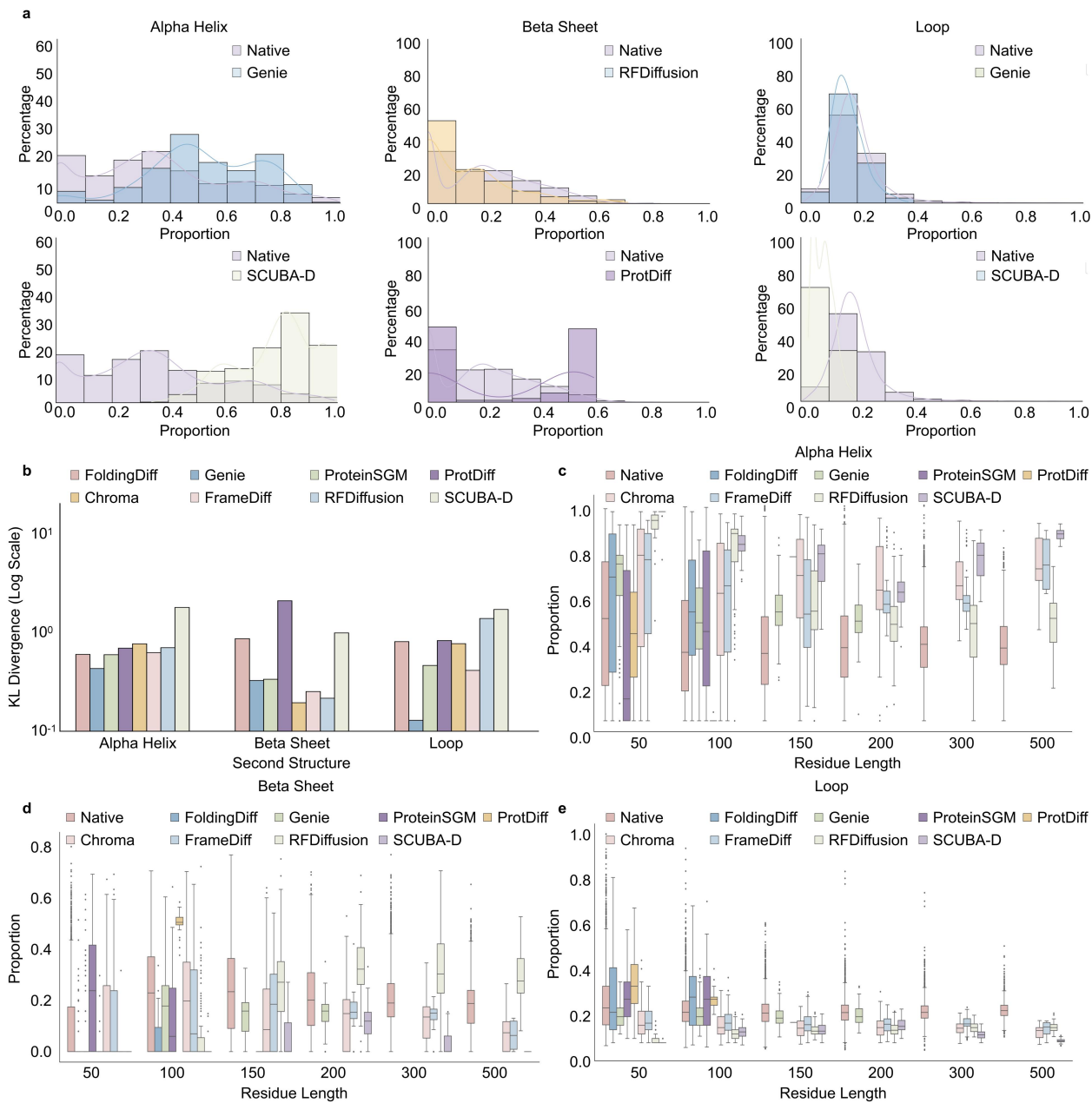

Fig. 5 a. Proportional distribution of different secondary structures in generated proteins compared to native proteins. b. KL divergence scores between the secondary structure distributions of generated and native proteins. c. Proportion of α-helix structures in generated and native proteins across different residue lengths. d. Proportion of β-sheet structures in generated and native proteins across different residue lengths. e. Proportion of loop regions in generated and native proteins across different residue lengths.

Statistical analysis reveals that, in terms of α-helix distribution, Genie-generated structures exhibit the highest similarity to native proteins. In contrast, SCUBA-D produces a slightly higher proportion of α-helices. For β-sheet distribution, Chroma-generated structures align most closely with distributions observed in native proteins. In contrast, ProtDiff shows the greatest deviation, characterized by the generation of proteins with either excessively low or high β-sheet content. Regarding loop region distribution, Genie again exhibits the highest similarity to the native proteins. In comparison, SCUBA-D exhibits the largest discrepancy, primarily due to an increased proportion of proteins with fewer loop regions.

Furthermore, we analyzed the impact of sequence length on the secondary structure distribution of the models (Fig. 5c, d, e). For α-helices, the proportion of generated proteins is generally higher than in native structures. ProteinSGM exhibits the closest α-helix proportion to native proteins when generating shorter structures, while RFdiffusion aligns best with natural distributions for longer structures. In β-sheet distribution, most models closely match native proteins for shorter structures; however, as residue length increases, the β-sheet proportion tends to decrease in most models, except for RFdiffusion. RFDiffusion produces a higher β-sheet content than native proteins for longer structures. In the loop region, shorter structures generated by most models show proportions comparable to native proteins. In contrast, for longer structures, the loop content is generally lower than that observed in native proteins.

## 4.5 Diversity

Diversity refers to the richness and variability of the generated proteins, reflecting a model's ability to produce proteins with distinct three-dimensional conformations. Here, we assess model diversity by clustering the generated structures using Foldseek clustering and hierarchical clustering[35]. Foldseek is a clustering method that integrates both structural and sequence information. However, since most generative models do not provide direct sequence information, we apply the 3Di Gotoh-Smith-Waterman alignment method to protein structures with a coverage threshold 0.9. To visually present the clustering results, we use t-SNE dimensionality reduction for visualization (Fig. 6a). Additionally, to quantitatively assess diversity, we count the number of clusters formed for each model (Fig. 6b). For hierarchical clustering, we first compute the TM-score and RMSD between generated structures, which serve as structural similarity and distance metrics, respectively. We then apply single-linkage clustering to categorize the generated structures. Specifically, we cluster model-generated proteins of different lengths using TM-score and RMSD as distance metrics, with a threshold of 0.5 for TM-score-based clustering and 2 for RMSD-based clustering. Furthermore, we count the number of clusters across different lengths (Fig. 6c) and visualize the heatmaps of the most and least populated clusters (Fig. 6d).

The Foldseek clustering results indicate that Chroma and RFDiffusion generate structures with greater diversity, whereas ProtDiff produces a more limited variety of structures (Fig. 6a). In the hierarchical clustering results, the number of clusters varies across proteins of different lengths. For shorter protein chains, FoldingDiff and Genie exhibit a higher number of structural categories (Fig. 6c). In contrast, for longer protein chains, Chroma, RFDiffusion, and SCUBA-D generate more diverse structures, with the number of clusters showing an increasing trend as protein length increases (Fig. 6c).

Combining these findings, Chroma, RFDiffusion and Genie demonstrates outstanding performance in diversity assessment. Additionally, the increase in residue length influences the structural diversity at a finer level, with RFDiffusion exhibiting greater diversity in longer protein sequences.

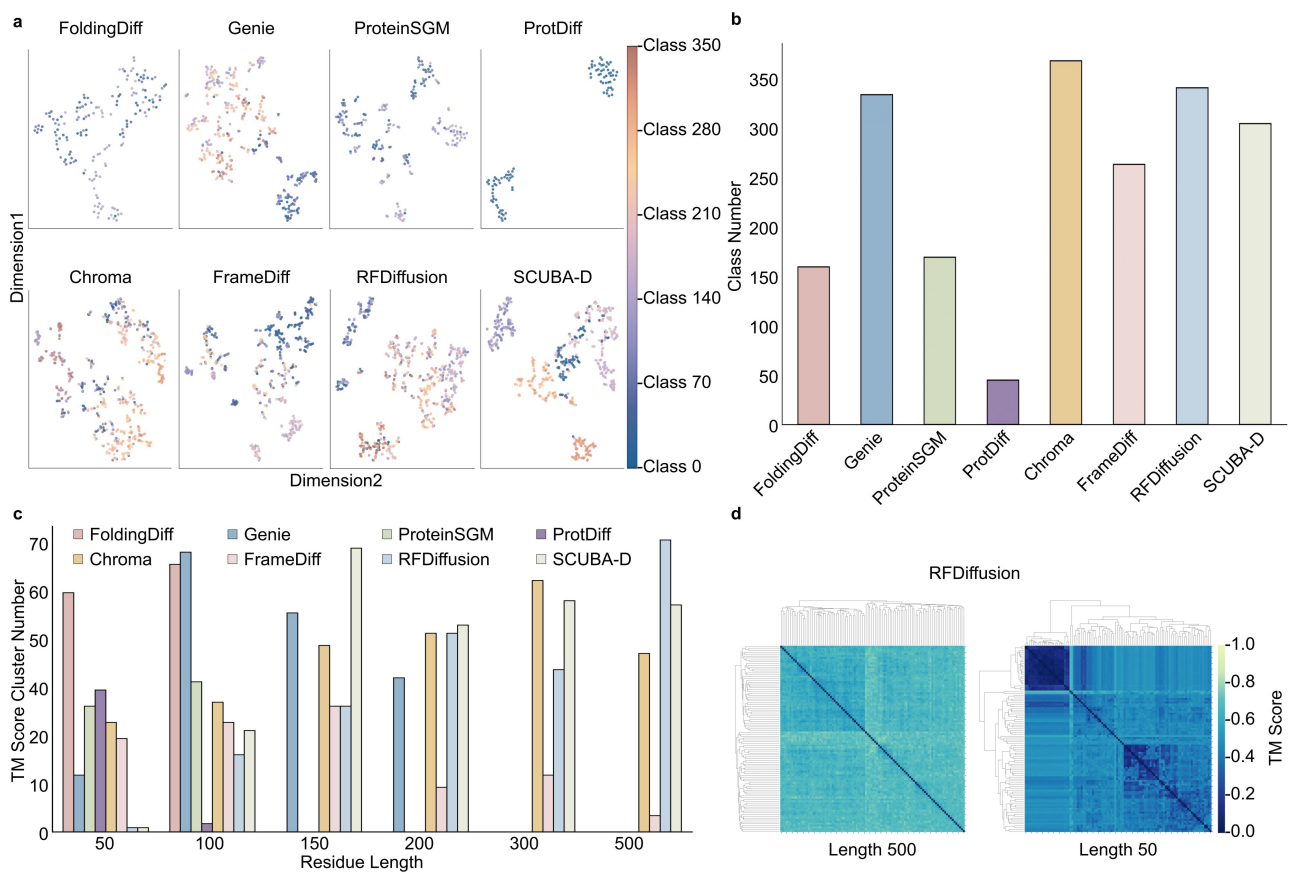

Fig. 6 a. T-SNE visualization of FoldSeek clustering results for generated proteins. b. Cluster count statistics of generated proteins based on FoldSeek clustering. c. Hierarchical clustering results using TM-score as the distance metric. d. Heatmap of hierarchical clustering results using TM-score as the distance metric.

## 4.6 Conditional Generation

Generating protein structures under specific constraints (Conditional generation) is a key metric for evaluating the ability of diffusion models. This capability reflects the potential of models in customized protein design. To comprehensively assess the conditional generation ability of each model, we define and analyze several representative conditional generation tasks (Table 4).

The evaluation results indicate that Chroma successfully completes all defined conditional generation tasks, demonstrating the broadest applicability across various research and application scenarios. RFdiffusion excels in spatially constrained design tasks but performs less effectively in abstract conditions (e.g., secondary structure or categorical constraints) and structure refinement tasks (e.g., Task E). ProtDiff, TaxDiff, and SCUBA-D show potential in specific application contexts but have limited overall functionality. Meanwhile, FoldingDiff, Genie, and FrameDiff currently do not support condition-based generation tasks.

The performance of different models in conditional generation tasks varies significantly. Chroma stands out for its diversity and flexibility, providing a powerful tool for complex protein design. In contrast, other models demonstrate potential in specific tasks but still exhibit functional limitations.

Table 4 Conditional Design Tasks

| Model | Binder Design | Symmetric Oligomer Design | Secondary Structure-constrained Design | Category-based Design | Structure Refinement | Partial Amino Acid Sequence-constrained Design |
|---|---|---|---|---|---|---|
| RFdiffusion | √ | √ | | | | √ |
| Chroma | √ | √ | √ | √ | √ | √ |
| EvoDiff | | | | √ | | √ |
| ProtDiff | √ | | | | | √ |
| TaxDiff | | | | √ | | |
| ProteinSGM | √ | | | | | √ |
| SCUBA-D | | | √ | | √ | |
| Foldingdiff | | | | | | |
| Genie | | | | | | |
| FrameDiff | | | | | | |

# 5 Successful Cases of Protein Design with Diffusion Models

Although various computational methods have been developed to evaluate model performance in protein design, biological experiments remain the gold standard for validating their effectiveness. This chapter focuses on protein design studies based on RFdiffusion, Chroma, and SCUBA-D. These studies not only confirm the validity of the models through experimental results but also highlight their strong potential in drug design.

RFdiffusion has demonstrated remarkable success in protein complex design, particularly in the targeted design of protein binders. Watson et al. fine-tuned the RFdiffusion model by incorporating hotspot residues at target protein-binding interfaces as input[8]. They designed a variety of protein binders, including influenza H1 hemagglutinin and interleukin-7 receptor. Using ProteinMPNN to generate amino acid sequences and AlphaFold2 for structural screening, they selected 95 candidate designs. Experimental validation through biolayer interferometry (BLI) achieved a success rate of 19%. Cryo-EM analysis confirmed that the highest-affinity influenza binder, HA_20, exhibited a structure highly consistent with the native complex, validating RFdiffusion's effectiveness in binder design. Similarly, Bennett et al. applied RFdiffusion to antibody complex design by fine-tuning the model on humanized VHH frameworks for multiple targets[37]. High-throughput screening and cryo-EM validation showed that the designed antibodies exhibited high affinity for their targets while maintaining structural accuracy. Additionally, Torres et al.

extended RFdiffusion to enable the flexible design of peptide binders, addressing limitations of traditional methods in designing helical-prone peptide interactions[38].

Chroma has been extensively applied in protein design, particularly in generating soluble proteins. John et al. utilized Chroma to generate 310 protein structures and conducted solubility experiments, revealing that most designed proteins exhibited excellent solubility[22]. Molecular biology experiments, including Western blotting and split-GFP assays, confirmed the solubility of multiple proteins. Further stability and secondary structure analyses demonstrated that the designed proteins exhibited remarkable thermal stability and structural integrity, aligning closely with the expected conformations.

Finally, the SCUBA-D model demonstrated outstanding performance in protein backbone design by integrating ABACUS-R for amino acid sequence generation, particularly in heme-binding proteins and Ras protein design. Liu et al. utilized SCUBA-D to design 30 heme-binding backbones and employed AlphaFold2 to filter the generated sequences[16]. Experimental results confirmed that the designed proteins effectively bound heme, with dissociation constants comparable to natural proteins. Additionally, in Ras protein design, among 360 backbone structures and 1,800 generated sequences, some designs exhibited strong binding affinity to natural Ras, further validating SCUBA-D's effectiveness in protein design.

These successful cases highlight the immense potential of diffusion-based models in protein design across various applications. From RFdiffusion's role in antibody and short peptide binder design, to Chroma's contributions to protein solubility and stability, and SCUBA-D's success in heme and Ras protein binder design, these studies collectively demonstrate the significant impact of diffusion models in diverse protein engineering tasks. While experimental validation remains essential for many designs, these successful cases provide valuable insights for future protein design research and underscore the promising real-world applications of diffusion models.

# 6 Discussion

With advancements in technology, researchers can not only discover proteins from nature but also create them. Artificial intelligence, particularly diffusion models, has significantly improved the success rate of de novo protein design. Diffusion models can be categorized into backbone generation models, sequence generation models and structure and sequence co-genration models Among them, structure design models employ different noise diffusion strategies depending on the data type, such as diffusion in feature maps, point clouds, or latent space. In recent years, structure design models like RFdiffusion have successfully designed various protein binders, some of which exhibit binding affinities surpassing existing proteins.

When evaluating existing protein design models, six key dimensions are considered: structural plausibility, designability, efficiency, diversity, novelty, and naturalness. Overall, RFdiffusion and Chroma exhibit the most balanced performance, while other models excel in specific aspects. The results indicate that different models excel in different dimensions (Fig. 2). FrameDiff shows superior structural plausibility and good naturalness, while RFDiffusion is known for its designability. Chroma excels in diversity and maintains naturalness, and FoldingDiff offers notable novelty with efficient performance.

Despite the significant advancements that diffusion models have brought to protein design, several limitations remain.

First, the *de novo* proteins are generated without considering their biocompatibility, making it challenging to ensure their safety when used for disease treatment. Second, protein functions in nature are often highly dynamic and frequently accompanied by structural changes. However, current models do not fully account for protein-ligand interactions or conformational flexibility during the design process. Additionally, while many models claim to be "end-to-end," they often require additional steps to generate amino acid sequences after predicting the protein structure. This not only increases computational time but also introduces potential noise that may compromise design quality. Lastly, existing protein design models have yet to directly optimize functional properties, such as activation or inhibition, which are critical for real-world applications. To achieve truly functional protein design, future research must rely more heavily on experimental validation and further explore structure-sequence co-generation strategies to enhance both efficiency and reliability.

# Supplement for The Dance of Atoms：De Novo Protein Design with Diffusion Model

Yujie Qin[1,2], Ming He[1,3], Changyong Yu[3], Ming Ni[1,*], Xian Liu[1,*], Xiaochen Bo[1,*]

[1]Department of Advanced & Interdisciplinary Biotechnology, Academy of Military Medical Sciences, Beijing, China

[2]School of Biological Science and Medical Engineering, Beihang University, Beijing, China

[3]College of Computer Science and Engineering, Northeastern University, Shenyang 110918, China

*Corresponding author(s). E-mail(s): niming@bmi.ac.cn, liux.bio@gmail.com, boxc@bmi.ac.cn ;
Contributing authors: qinyujie@buaa.edu.cn;
[†]These authors contributed equally to this work.

# Supplementary Figure 1

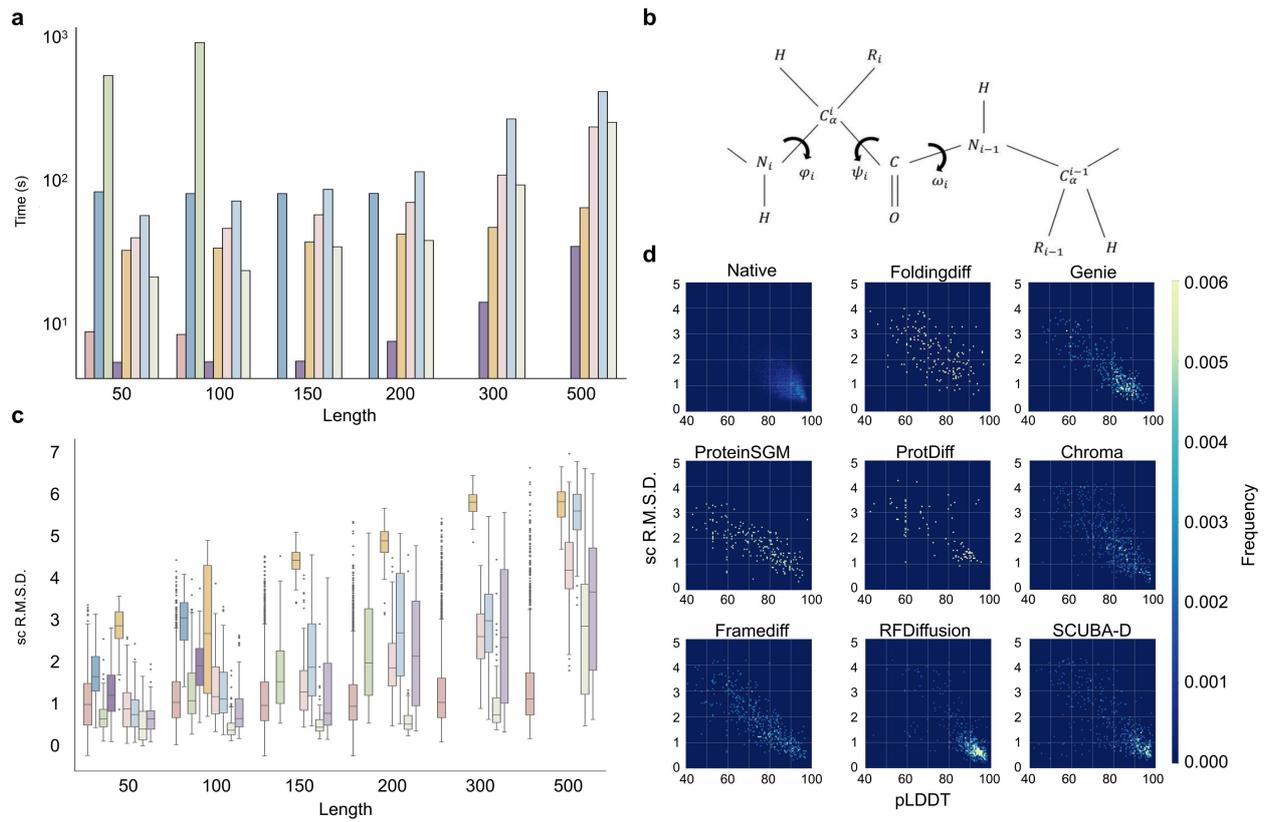

Supplementary Figure 1: Model Efficiency and Designability. a) Average generation time per structure for the models. b) Schematic diagram of protein dihedral angles. c) Self-consistent R.M.S.D. scores of model-generated structures at different lengths. d) Relationship between sc-R.M.S.D. and pLDDT for native and generated proteins.

# Supplementary Figure 2

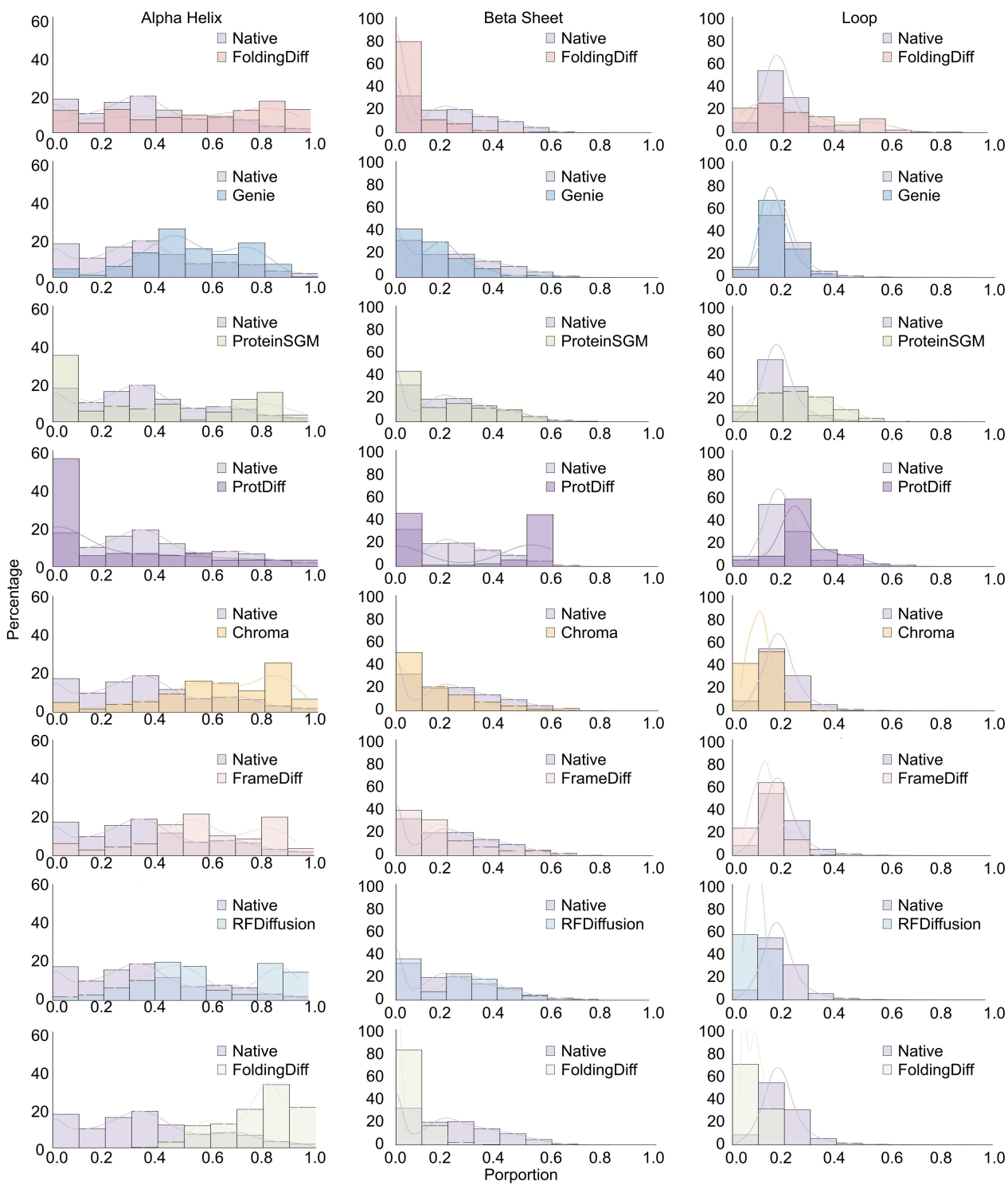

Supplementary Figure 2: Proportional distribution of different secondary structures in generated proteins compared to native proteins.

# Supplementary Figure 3

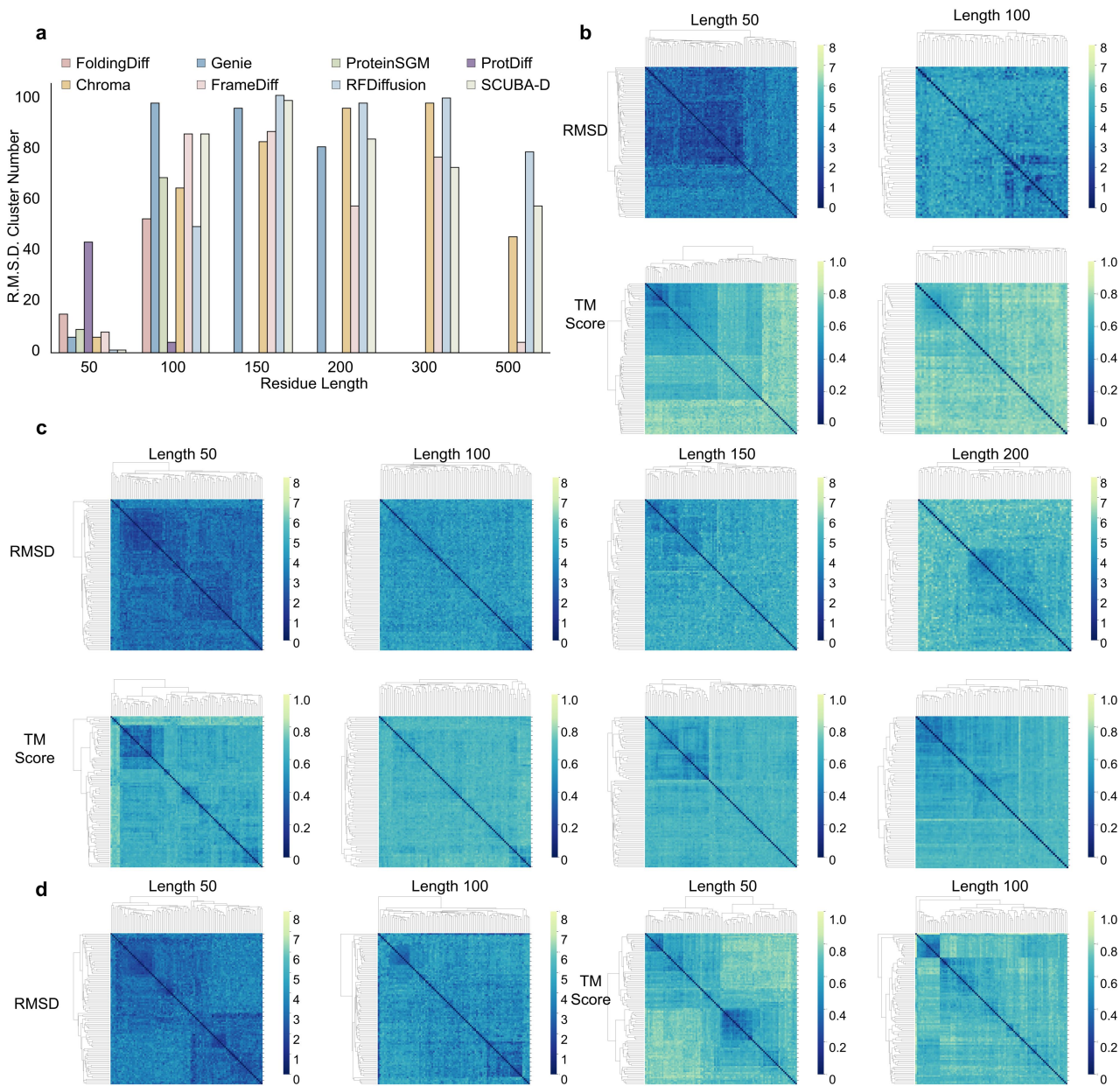

Supplementary Figure 3: Hierarchical clustering results using R.M.S.D. as the distance metric. a) Hierarchical clustering results using R.M.S.D. as the distance metric. b) Heatmap of FoldingDiff hierarchical clustering results using R.M.S.D. as the distance metric. c) Heatmap of Genie hierarchical clustering results using R.M.S.D. as the distance metric. d) Heatmap of ProteinSGM hierarchical clustering results using R.M.S.D. as the distance metric.

# Supplementary Figure 4

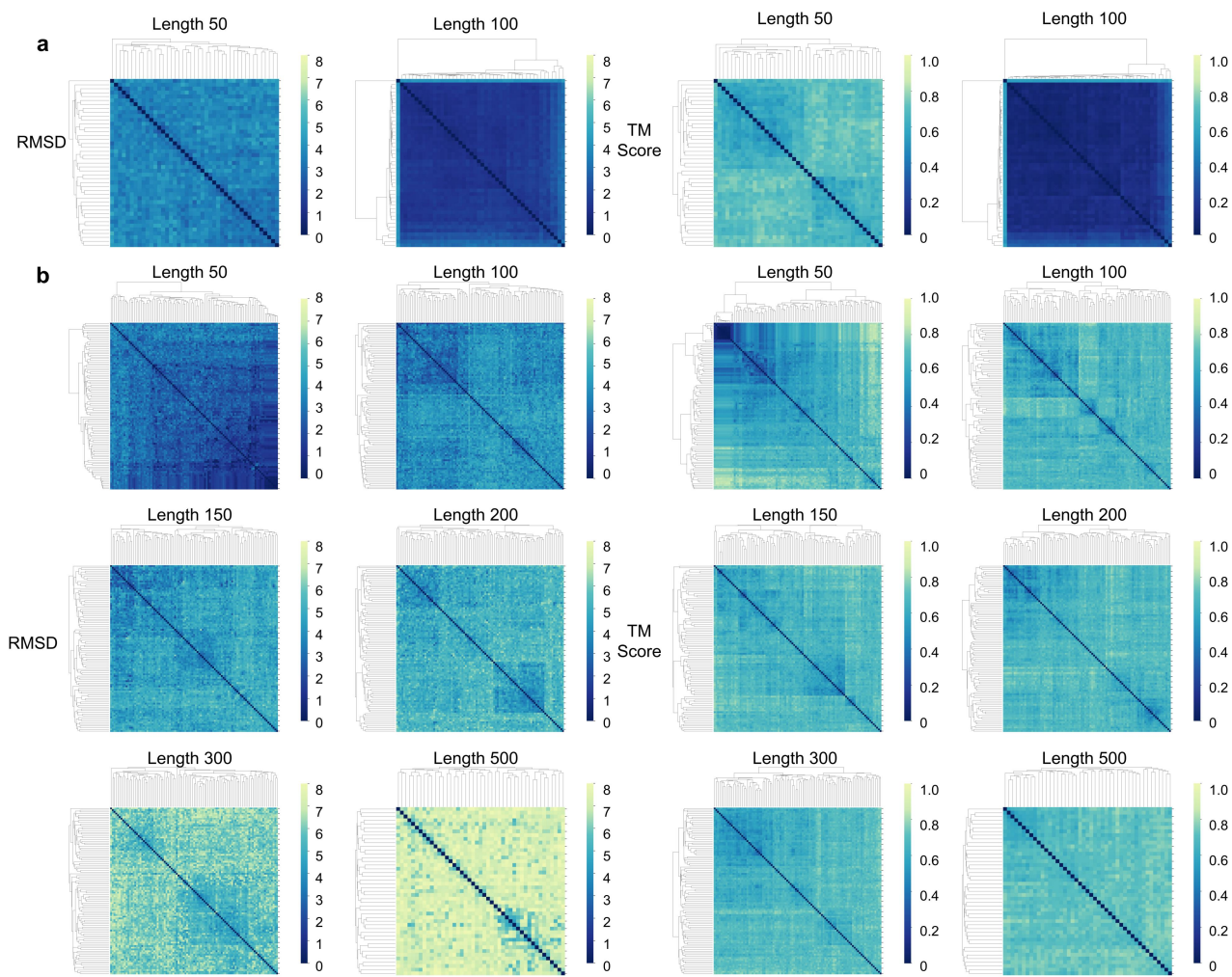

Supplementary Figure 4: Hierarchical clustering results using R.M.S.D. as the distance metric. a) Heatmap of ProtDiff hierarchical clustering results using R.M.S.D. as the distance metric. b) Heatmap of Chroma hierarchical clustering results using R.M.S.D. as the distance metric.

# Supplementary Figure 5

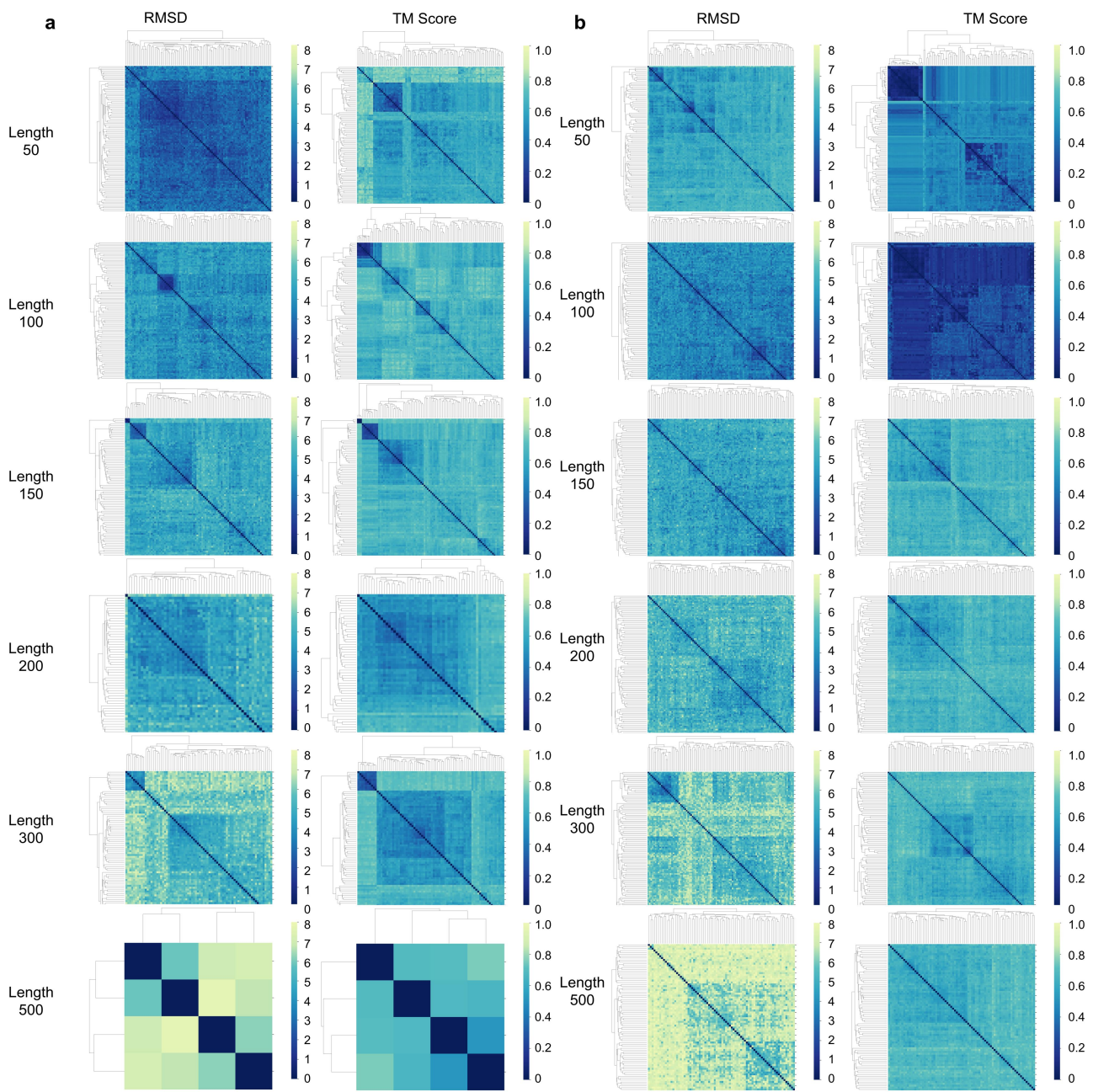

Supplementary Figure 5: Hierarchical clustering results using R.M.S.D. as the distance metric. a) Heatmap of FrameDiff hierarchical clustering results using R.M.S.D. as the distance metric. b) Heatmap of RFDiffusion hierarchical clustering results using R.M.S.D. as the distance metric.

# Supplementary Figure 6

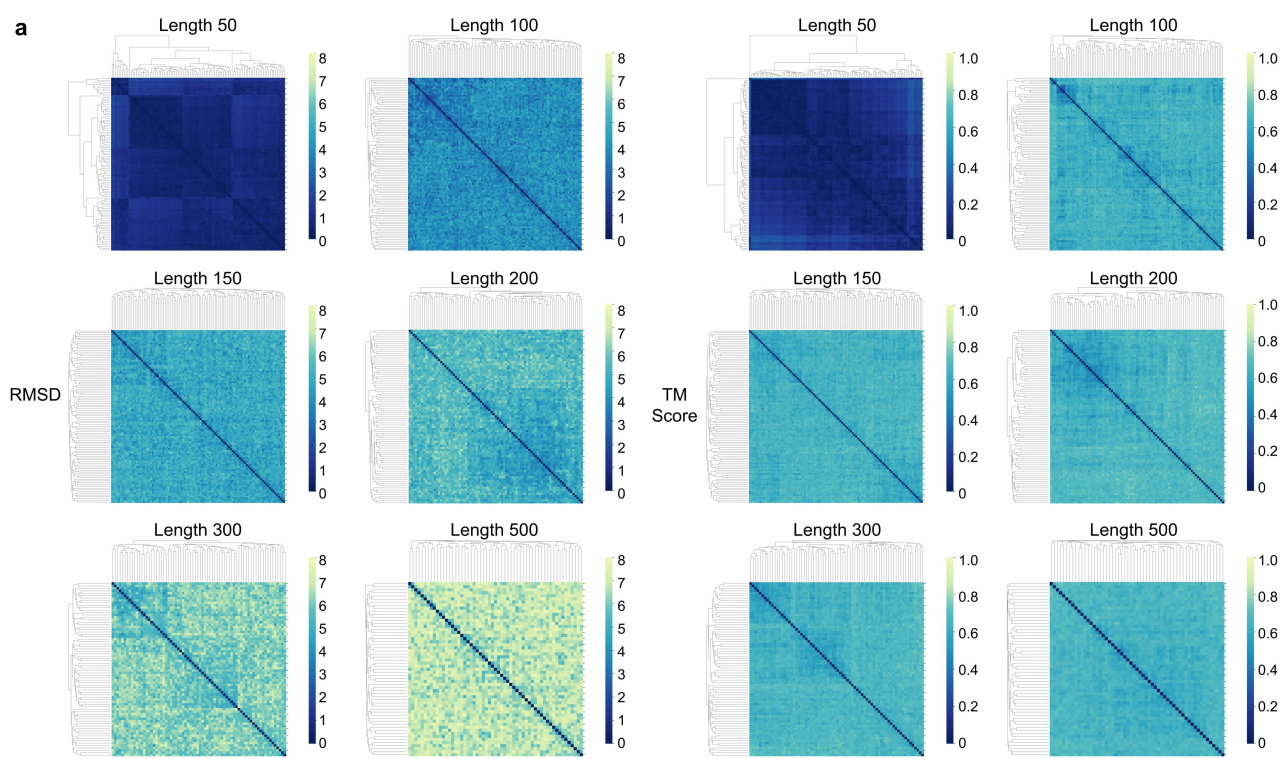

Supplementary Figure 6: Hierarchical clustering results using R.M.S.D. as the distance metric. a) Heatmap of SCUBA-D hierarchical clustering results using R.M.S.D. as the distance metric.